\documentclass[conference]{IEEEtran}
\IEEEoverridecommandlockouts
\usepackage{algorithm}
\usepackage{algpseudocode}
\usepackage{tabularray}
\UseTblrLibrary{booktabs}
\usepackage{graphicx}
\usepackage{amsmath,amssymb,amsfonts}
\usepackage{pifont}% http://ctan.org/pkg/pifont
\usepackage[utf8]{inputenc}
\usepackage[numbers,sort&compress]{natbib}
\usepackage[a4paper, total={184mm,239mm}]{geometry}
\def\BibTeX{{\rm B\kern-.05em{\sc i\kern-.025em b}\kern-.08em
    T\kern-.1667em\lower.7ex\hbox{E}\kern-.125emX}}

\newcommand{\cmark}{\ding{51}}%
\newcommand{\xmark}{\ding{55}}%
\newcommand{\bcmark}{\ding{52}}%

\begin{document}

\title{COMPASS: A Compiler Framework for Resource-Constrained Crossbar-Array Based In-Memory Deep Learning Accelerators
}

\author{
\IEEEauthorblockN{Jihoon Park$^*$, Jeongin Choe$^*$, Dohyun Kim, Jae-Joon Kim}
\IEEEauthorblockA{\textit{Seoul National University}, Seoul, South Korea}
}
% \author{
% \IEEEauthorblockN{Anonymous Authors}
% \vspace*{0mm}
% }
\maketitle

% \renewcommand{\shortauthors}{Park et al.}

\begin{abstract}
  Recently, crossbar array based in-memory accelerators have been gaining interest due to their high throughput and energy efficiency. While software and compiler support for the in-memory accelerators has also been introduced, they are currently limited to the case where all weights are assumed to be on-chip. This limitation becomes apparent with the significantly increasing network sizes compared to the in-memory footprint. %As it is not the case with largely increasing network sizes compared to in-memory footprint,
  Weight replacement schemes are essential to address this issue. We propose \textit{COMPASS}, a compiler framework for resource-constrained crossbar-based processing-in-memory (PIM) deep neural network (DNN) accelerators. \textit{COMPASS} is specially targeted for networks that exceed the capacity of PIM crossbar arrays, necessitating access to external memories. We propose an algorithm to determine the optimal partitioning that divides the layers so that each partition can be accelerated on chip. Our scheme takes into account the data dependence between layers, core utilization, and the number of write instructions to minimize latency, memory accesses, and improve energy efficiency. Simulation results demonstrate that \textit{COMPASS} can accommodate much more networks using a minimal memory footprint, while improving throughput by 1.78X and providing 1.28X savings in energy-delay product (EDP) over baseline partitioning methods.
\end{abstract}

% \keywords{PIM, Accelerator, Neural Network, CNN, In-memory Computing, Compiler, Layer pipelining }

\maketitle
\def\thefootnote{*}\footnotetext{These authors contributed equally to this work}\def\thefootnote{\arabic{footnote}}

%%%%%%%%%%%%%%%%%%%%%%%%%%%%%%%%%%%%%%%%%%%%%%%%%%%%%%%%%%%%%%%%%%%%%%%%%%%%%
%%%%%%%%%%%%%%%%%%%%%%%%%%%%%%%%%%%%%%%%%%%%%%%%%%%%%%%%%%%%%%%%%%%%%%%%%%%%%
\section{Introduction}
Rapid evolution of Deep Neural Networks (DNNs) has fueled the demand for advanced computing architectures and accelerators to meet growing computational requirements. As DNNs continue to scale in complexity and size, conventional computing architectures, primarily Von Neumann, face significant challenges in coping with the escalating demand for computation and memory bandwidth. The traditional separation of processing and memory, inherent in Von Neumann architectures, introduces substantial inefficiencies, particularly in data movement and energy consumption. Recognizing these challenges, attention to Processing-In-Memory (PIM) architectures has been rapidly increasing as an alternative to traditional architecture~\cite{shafiee2016isaac,song2017pipelayer,sun2023pimcomp,qiao2018atomlayer}.
%Within the PIM paradigm, SRAM-based PIM accelerators have garnered considerable interest~\cite{valavi201964,jia202115,yin2021pimca,ueyoshi2022diana, tu202228nm}. The appeal of SRAM lies not only in its versatility but also in its maturity in process technology. However, current SRAM-based PIM accelerators need tailored weight mapping and execution schemes due to the lack of compiler support. 
Along with the architectural interest, IMC are being implemented in various technologies such as SRAM~\cite{jia202115,yin2021pimca}, ReRAM~\cite{xue202116,huang2023nonvolatile,hung20228}, MRAM~\cite{roy2024compute,cai202333,deaville2024fully}, etc. Recent prototype chips integrate multiple macros on chip to verify the effectiveness of PIM architectures. However, these prototypes still suffer from limited IMC memory capacity, regardless of which technology it uses, varying from a few hundred KBs to a few MBs and do not support networks that exceeds the size of the IMC footprint at all. Some includes tailored weight mapping and execution schemes but these approaches are not general with lack of compiler support. Existing compilers for digital neural processing units (NPUs) are not suitable for PIMs as they do not consider the PIM's inherent parallel MVM operation capability. PIM-aware compilers like PUMA~\cite{ankit2019puma} and PIMCOMP~\cite{sun2023pimcomp} have their primary focus on mapping all the weights on chip, but it is not possible to map large networks on chip when PIM memory footprint is constrained to tens of MBs at most.

\begin{figure}[t]
    \centering
    \includegraphics[width=0.95\linewidth]{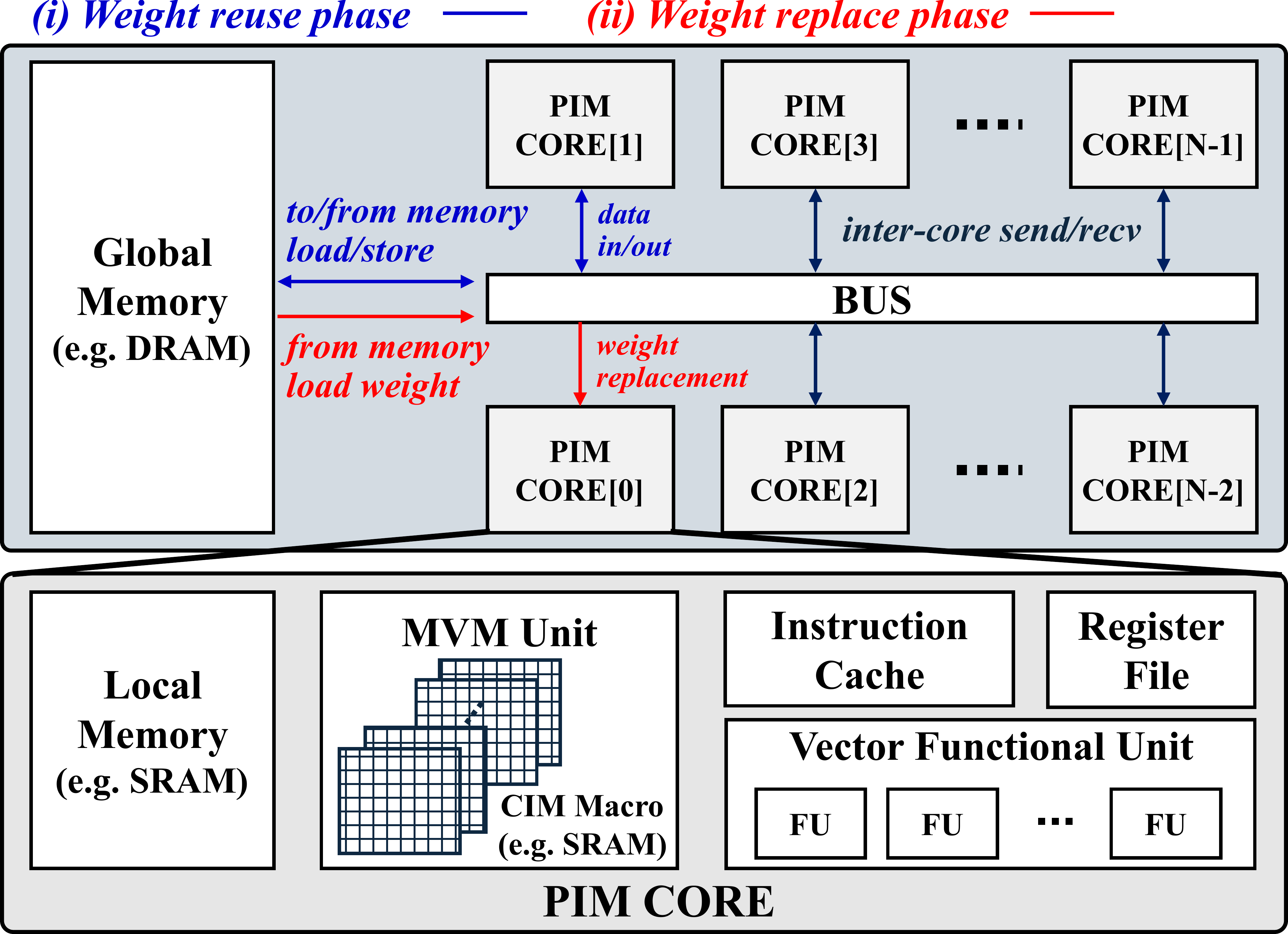}
    \vspace*{-2mm}
    \caption{In-memory DNN accelerator architecture with weight replacement}
    \label{fig:architecture_overview}
    \vspace*{-1mm}
\end{figure}

To address the issue, we introduce a novel compiler framework, \textit{COMPASS}, designed for resource-constrained crossbar-based in-memory DNN accelerators. Our framework intelligently partitions the network into smaller and manageable units, optimizing on-chip resource utilization while maintaining balance across layers and partitions. Through this approach, we provide a pragmatic solution to the challenges posed by mapping large DNNs on PIM design with limited resources. Our contributions can be summarized as follows:
\begin{itemize}
\item We introduce a compiler framework for in-memory computing, which can support large DNN models when the required weight memory exceeds the CIM memory footprint, requiring communication with external memory.
\item We propose a network partitioning method for general crossbar-array based in-memory accelerators by supporting weight reloading and multi-endpoint dependency checks between each partition.
\item We propose a genetic algorithm (GA) to find the optimal partitioning and a novel fitness function that optimizes for high throughput and low energy-delay product (EDP).
\end{itemize}

%%%%%%%%%%%%%%%%%%%%%%%%%%%%%%%%%%%%%%%%%%%%%%%%%%%%%%%%%%%%%%%%%%%%%%%%%%%%%
%%%%%%%%%%%%%%%%%%%%%%%%%%%%%%%%%%%%%%%%%%%%%%%%%%%%%%%%%%%%%%%%%%%%%%%%%%%%%
\section{PIM accelerator with model partitioning}
\subsection{Weight Replacement}
\label{ssec:weight_replacement}

\begin{figure}[t]
    \centering
    \includegraphics[width=1.0\linewidth]{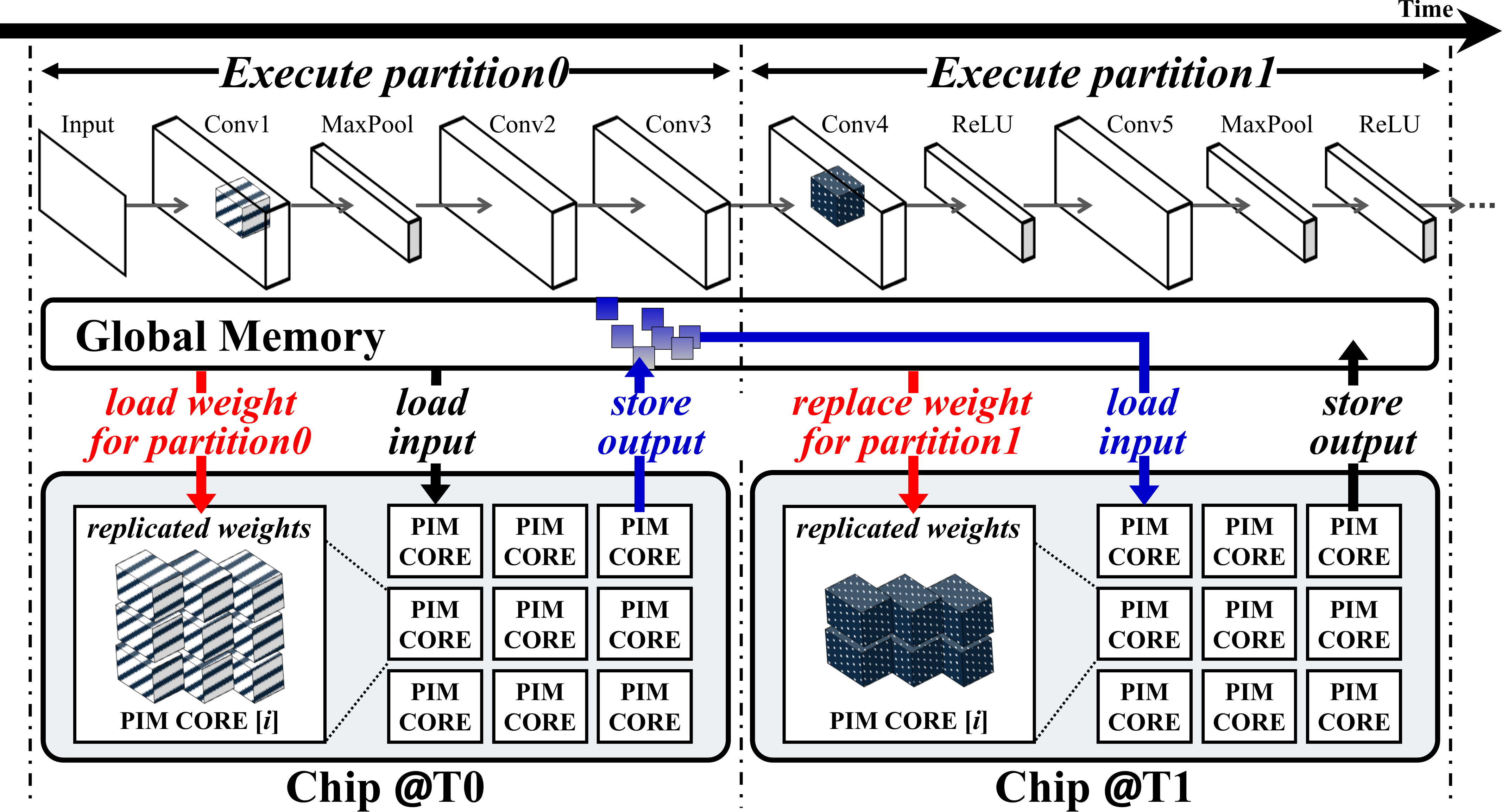}
    \vspace*{-5mm}
    %\caption{Partitioned model execution}
    \caption{Partitioned model execution. At T0, the first partition runs with weights loaded into PIM memory, inputs processed, and outputs stored in the global memory. The stored output becomes the input for the next partition at T1.}
    \label{fig:partition_overview}
\end{figure}

We choose to use the abstract in-memory DNN accelerator architecture described in \cite{sun2023pimcomp} as a PIM architecture template. Similar to previous works \cite{shafiee2016isaac, song2017pipelayer, ankit2019puma}, the architecture defines a Macro-Core-Chip hierarchy as shown in Fig. \ref{fig:architecture_overview}. The chip consists of multiple PIM cores and a global memory, where each PIM core is connected via on-chip interconnect. Each core contains a matrix unit, a vector processing unit, an instruction memory for core control, and local registers and memory for storing intermediate results and activation features. The matrix unit consists of multiple crossbar-based CIM macros which perform matrix-vector multiplications in an efficient manner.

To address cases where a large DNN model does not fit within the CIM memory footprint, we introduce a weight replacement capability, as depicted in color in Fig. \ref{fig:architecture_overview}. During the weight reuse phase, a partition of the model is mapped and executed on chip, loading input and storing output in the process. The core then transitions to a weight replace phase, where weights are loaded from global memory and broadcast to the crossbars for writing.
%Then, the core transitions into a weight replace phase. In this phase, the core loads the weights from the global memory to its local core and then broadcasts the weights to the crossbars where the weight should be written. 
When the core enters a computation phase again, new inputs are loaded from global memory, and the generated results are stored back upon exiting the state.

%%%%%%%%%%%%%%%%%%%%%%%%%%%%%%%%%%%%%%%%%%%%%%%%%%%%%%%%%%%%%%%%%%%%%%%%%%%%%
%%%%%%%%%%%%%%%%%%%%%%%%%%%%%%%%%%%%%%%%%%%%%%%%%%%%%%%%%%%%%%%%%%%%%%%%%%%%%
\subsection{Partitioned Model Execution}
\label{ssec:partitioned_exec}

\begin{figure}[t]
    \centering
    \includegraphics[width=1.0\linewidth]{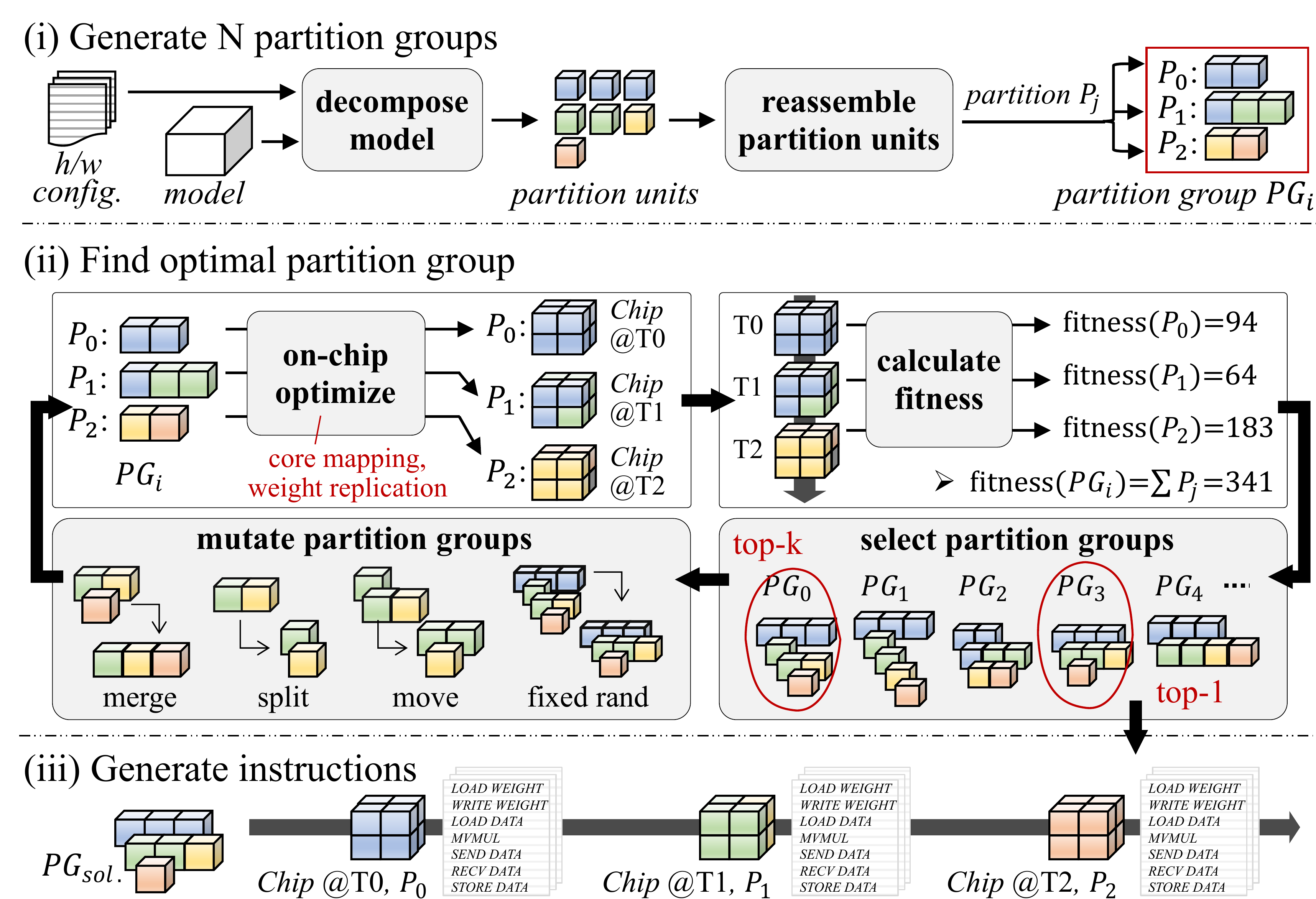}
    \vspace*{-5mm}
    \caption{\textit{COMPASS} compiler framework overview}
    \label{fig:compiler_framework}
\end{figure}

%Fig.~\ref{fig:partition_overview} shows the abstract view of how the cores are mapped based on the compiler's output. 
%The model, represented in a Directed Acyclic Graph (DAG), is split into multiple partitions, where each partition contains a portion of the DAG. Each node within the partition is replicated and mapped to the crossbars inside the chip. 
Fig.~\ref{fig:partition_overview} shows the abstract view of the partitioned model execution. 
The execution within the partition is conducted in a pipelined manner, treating each partition as a model fully mapped on chip. The execution of different partitions is performed sequentially, with weights being replaced between each execution. Cores mapped to earlier layers complete execution and start weight replacement faster, enhancing the effective utilization of global memory bandwidth.

Another critical factor we consider is the weight replication. Weight replication is a widely known scheme to replicate the weights of the layers prior to pooling or striding to balance the throughput between pipelined layers. However, previous works~\cite{song2017pipelayer,sun2023pimcomp} assume that a DNN model fits entirely on a chip and they rather do a simple optimization to determine how to allocate the remaining in-memory computing cells for additional replication of layers. Considering weight replacement together with weight replication becomes a more complex joint optimization problem in which partitioning the model and replicating layers within the partition should both be considered for optimal performance. %However, previous works do not determine the replication scheme when a DNN model does not entirely fit on a chip. When large DNN model is divided up into multiple partitions, it is important to determine whether to use crossbars to replicate existing layers or to assign them to more layers for deeper pipelining. 
%Determining which portion of the model to load and how many replications to have is crucial as it significantly affects the overall performance of the execution. 
%While replicating the weights on crossbar arrays is essential for balancing the throughput, it leaves less room to map remaining weights, thereby limiting the number of layers the chip can handle with its on-chip weight. In addition, high replication numbers can result in more partitions and increase the number of global memory accesses. Last but not least, splitting a layer across different cores or phases is beneficial for mapping utilization, but it can incur more frequent inter-core communication and global memory accesses. 

Additionally, we propose to execute each partition in a batched manner, where weight parameters of each partition are loaded and reused until a batch of input features are processed and saved to DRAM. Then, the weights are replaced for the next partition execution. Increasing the batch size improves the throughput and per-inference energy consumption. On the other hand, each sample has to wait for other samples in the same batch to finish before starting execution of its next partition, thereby increasing the end-to-end latency of the total execution. Therefore, the batch size should be kept relatively small to balance the throughput and the end-to-end inference latency.

%%%%%%%%%%%%%%%%%%%%%%%%%%%%%%%%%%%%%%%%%%%%%%%%%%%%%%%%%%%%%%%%%%%%%%%%%%%%%
%%%%%%%%%%%%%%%%%%%%%%%%%%%%%%%%%%%%%%%%%%%%%%%%%%%%%%%%%%%%%%%%%%%%%%%%%%%%%
\section{compiler Framework}
\subsection{Overview of COMPASS Framework}

%FIG: \textit{COMPASS} overview\newline
Fig.~\ref{fig:compiler_framework} shows the proposed \textit{COMPASS} compiler framework which consists of three major components: partition generator, partition optimizer, and a scheduler. A user-specified hardware configuration (e.g. crossbar attributes, number of crossbars, core size, interconnect specification) and the network model is provided to the partition generator. Subsequently, the partition generator divides the model into the smallest units for core mapping (partition unit) and then recombines these units to generate initial model partitions (partition group). The partition optimizer uses the initial partition groups to estimate their performance considering the weight replication and core mapping and iteratively optimizes the partition groups. We introduce the \textit{COMPASS} algorithm which is a GA algorithm to select and mutate the partition groups. After multiple generations, it identifies the optimal partition. Finally, the scheduler generates necessary instructions for model execution on each core, including the weight writes and activation load/store instructions between partitions as described in Sec.~\ref{ssec:weight_replacement}.

%%%%%%%%%%%%%%%%%%%%%%%%%%%%%%%%%%%%%%%%%%%%%%%%%%%%%%%%%%%%%%%%%%%%%%%%%%%%%
%%%%%%%%%%%%%%%%%%%%%%%%%%%%%%%%%%%%%%%%%%%%%%%%%%%%%%%%%%%%%%%%%%%%%%%%%%%%%

\begin{figure}[t]
    \centering
    \includegraphics[width=1.0\linewidth]{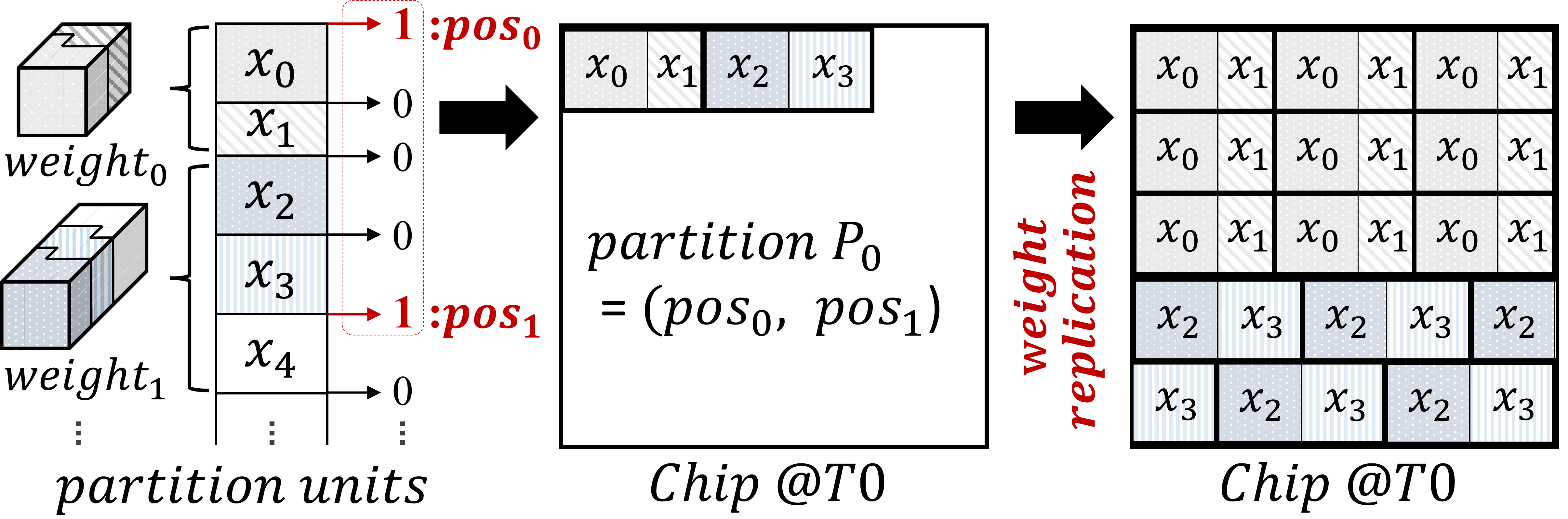}
    \vspace*{-5mm}
    \caption{Model decomposition and partition generation}
    \label{fig:partitioning}
\end{figure}

\subsection{Partition Generation}
\label{ssec:partition_gen}

Fig.~\ref{fig:partitioning} illustrates the process of model decomposition and partition generation.
First, weight matrices are divided along the output dimension into partition units $x_i$'s. Each unit is sized to fit within the IMC memory footprint of a single core, serving as a minimum granularity for partitioning.
Then, a span of consecutive partition units represented by a pair of position, $\{x_i \mid pos_0 \leq i < pos_1\}$, are grouped into the same partition $P_0$.
As the last step, weights are replicated in each partition as necessary.
% Each kernel is divided into sub-kernels called partition units $x_i$ which serve as the minimum granularity for partitioning. Then, a span of consecutive partition units represented by a pair of positions, $\{x_i \mid pos_0 \leq i < pos_1\}$, are placed in the same partition $P_0$. As the last step, weights are replicated in each partition as necessary.
%the replication of the partition units are made in each partition. 
The following conditions should be met for valid partitioning and replicating:

\begin{enumerate}
\item [1.] A partition unit cannot be of size bigger than the in-memory footprint of a single core. 
\item [2.] The partition units that originate from a single kernel share their replication counts. 
\item [3.] The total size of the replicated units cannot exceed the chip memory constraint. 
\end{enumerate}

Through this procedure, the Conv/Linear layers can be flexibly assigned to their respective partition according to the model and hardware constraint. Early layers with small kernel sizes can be mapped together inside a single partition with replication, while later layers with bigger kernel size can be split into multiple partitions. A validity map is constructed for efficient partition selection. The layers that cannot be mapped on crossbar arrays are dealt after the partition is generated.

%%%%%%%%%%%%%%%%%%%%%%%%%%%%%%%%%%%%%%%%%%%%%%%%%%%%%%%%%%%%%%%%%%%%%%%%%%%%%
\subsubsection{Validity Map}
\begin{figure}[t]
    \centering
    \includegraphics[width=0.9\linewidth]{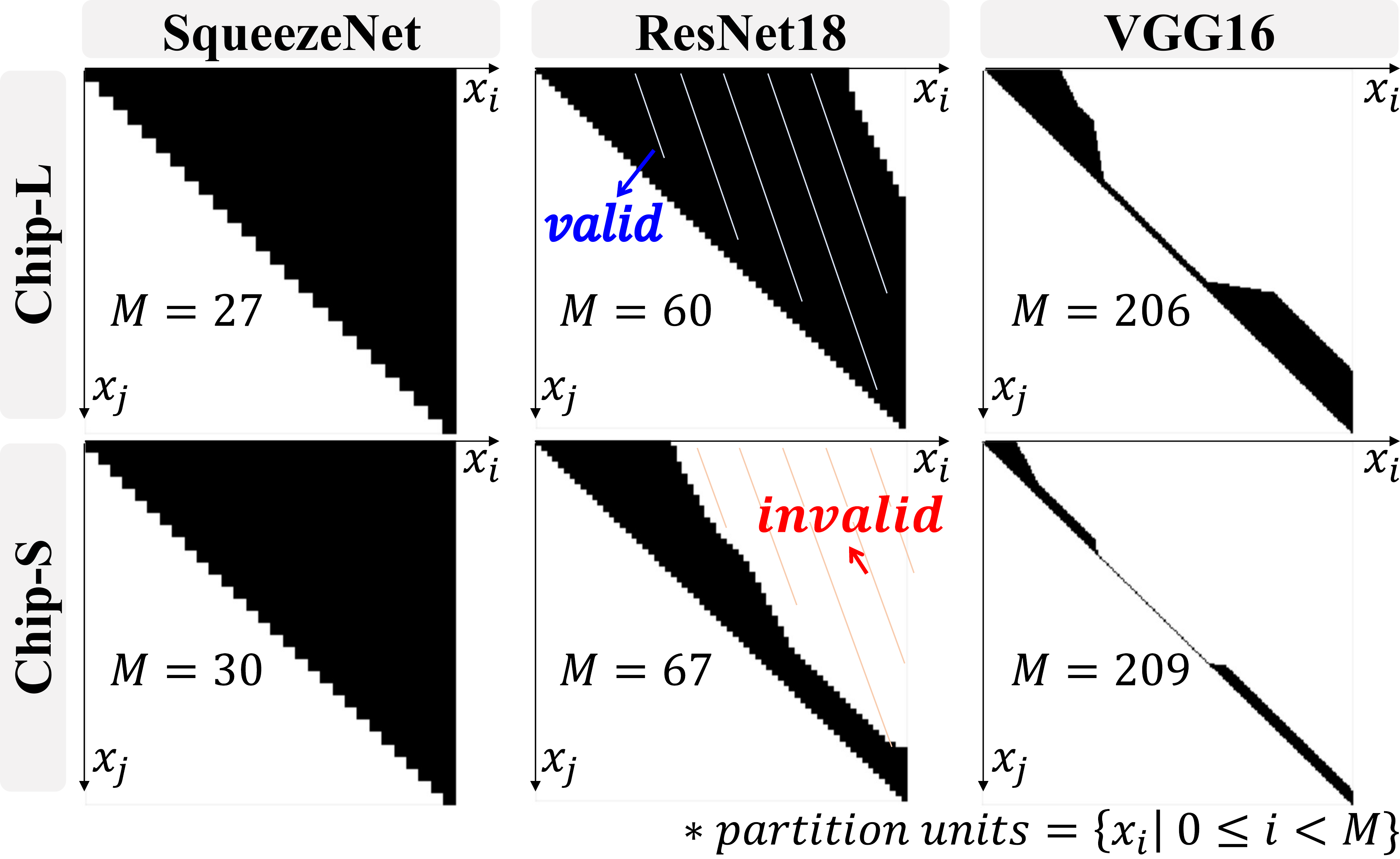}
    \vspace*{-3mm}
    \caption{Partition validity map. Chip-S and Chip-L represent the small and large chip configurations, as detailed in Table~\ref{tab:hw_config}. The models increase in size from SqueezeNet to VGG16, with details provided in Table~\ref{tab:network}.}
    \label{fig:validity_map}
\end{figure}

If partition positions are selected randomly, the likelihood of producing a valid result becomes low and multiple iterations are required to find a valid solution, especially with a large model size and a small in-memory computing cell capacity.  Therefore, instead of randomly selecting positions to generate partitions, we pre-calculate a validity map which marks the possible end position when a starting position for a partition is given. Using the validity map, we can ensure that every partition is generated within the chip's constraints. Fig.~\ref{fig:validity_map} shows the validity map for different model and chip sizes, indicating whether a partition defined by two positions ($x_i$, $x_j$) is valid. $M$ represents the number of partition units after model decomposition in each case. We iteratively select partition positions within the valid range, taking into account the positions previously selected. Note that with a bigger number of weight parameters and a smaller in-memory computing cell capacity (towards the lower right in Fig.~\ref{fig:validity_map}), the invalid portion of the validity map becomes larger.

%하나의 파티션은 연속한 파티션 유닛의 배열이므로 시작 포인트와 엔드 포인트의 두 값으로 정의 될 수 있으며, 따라서 When there are $M$ partition units for the model, there are $M+1$(M+1)C2 possible position pairs to select from.For all partitions created by the position pairs, 이것은 whether the required resource does not exceed the chip constraint인지 체크될 필요가 있고 when validation fails 시, regenerates the positions 할 필요가 있다. 그러나, 모델사이즈가 커지고 하드웨어가 작아지면 M이 증가하며 position regeneration 비용이 비싸지므로 파티션 유닛의 ccululated crossbar resource를 statically 계산하여 MxM validity map을 저렴하게 구현하여, 가능한 파티셔닝 포지션이 이 validity map에 의해 프루닝 될 수 있도록 하였다. Fig.~\ref{fig:validity_map} shows the case where possible(?) partitions is passes the validation 3 모델과 두가지 타입의 하드웨어에 따라서. 파티션 제너레이터는 이 밸리디티맵을 기반으로 x_0부터의 가능한 end point를 random하게 순차적으로 생성해 나가다,마지막 파티션 유닛에 도달할때까지. 

%%%%%%%%%%%%%%%%%%%%%%%%%%%%%%%%%%%%%%%%%%%%%%%%%%%%%%%%%%%%%%%%%%%%%%%%%%%%%
\subsubsection{Non-crossbar-mapped Layers}
After the partitions for layers that are to be mapped on crossbar are determined, other layers are taken into account. We construct a directed acyclic graph (DAG) for the decomposed model, with partition units serving as nodes. As the remaining non-crossbar-mappable layers such as batch normalization and activation layers are closely coupled with the prior Conv/Linear layer, we traverse back the layer dependence graph to place the layers in the same partition. 

%%%%%%%%%%%%%%%%%%%%%%%%%%%%%%%%%%%%%%%%%%%%%%%%%%%%%%%%%%%%%%%%%%%%%%%%%%%%%
\subsubsection{Memory Access Management}
During an execution of a partitioned graph, the entry nodes require load from the global memory, and the exit nodes have to store the intermediate features back to the global memory for next partition execution. Thus these nodes are marked with load/store attributes along with the respective data transfer sizes. Unlike a fully on-chip model, which has a single entry and exit node, each partition can have multiple entry and exit nodes. For example, a ResNet model with a residual connection that is not fully contained within a partition would have multiple exit nodes. The load/store attributes are used to consider DRAM access latency in performance estimation, memory allocation during instruction scheduling. 

\subsection{COMPASS Algorithm}

Algorithm~\ref{alg:algorithm} outlines the steps of the \textit{COMPASS} algorithm. Each gene represents a partition and each chromosome represents a partition group. The partition generator generates a predefined number of initial partition groups, denoted as $\Pi_0$. In each generation $g$, a predefined ratio of population is kept ($n_{sel}$) and then mutated to generate the remaining population ($n_{mut}$) based on their fitness (Sec.~\ref{sssec:pg_fit}). For partition groups selected for mutation, score for each partition is calculated (Sec.~\ref{sssec:p_score}) and the group is mutated by one of the four mutation schemes (Sec.~\ref{sssec:mutation}). At the end of the generation ($g = G$), the final partition group $\phi$ is selected based on its fitness.

%% https://tug.ctan.org/macros/latex/contrib/algorithmicx/algorithmicx.pdf
\begin{algorithm}[t]
\caption{\text{ \textit{COMPASS} algorithm($G, N, n_{sel}, n_{mut}$)}}
\label{alg:algorithm}
\begin{algorithmic}[1]
\State $\Pi_0 \gets generatePGs(N)$  
\For{$g \gets 0, G$}
  \ForAll {$PG_i \in \Pi_g$}
    \State $PGF_i \gets \Call{CalculatePGFitness}{PG_i}$
  \EndFor
  \State $\Pi_g \gets sort_{asc}(\Pi_g; \text{PGF})$
  \State $\Pi_{sel} \gets takeFirstN(\Pi_g, n_{sel})$
  \State $\Pi_{mut} \gets takeRandomN(\Pi_{sel}, n_{mut})$  
  \ForAll {$PG_j \in \Pi_{mut}$}
    \ForAll {$P_k \in PG_j$}
      \State $R_k \gets \Call{CalculatePartitionScore}{P_k}$
    \EndFor
    \State $PG_j \gets sort_{asc}(PG_j; \text{R})$
    \State $P_{mut} \gets takeLastN(PG_j, 1)$
    \State $PG_j \gets mutatePG(PG_j, P_{mut})$
  \EndFor
  \State $\Pi_{g+1} \gets \Pi_{sel} \cup \Pi_{mut}$ 
\EndFor
\State $\Pi_G \gets sort_{asc}(\Pi_G; \text{PGF})$
\State $\phi \gets takeFirstN(\Pi_G, 1)$
\State \textbf{return} $\phi$

\end{algorithmic}
\end{algorithm}
%FIG or PSEUDO CODE: partition optimizer algorithm flow\newline

%%%%%%%%%%%%%%%%%%%%%%%%%%%%%%%%%%%%%%%%%%%%%%%%%%%%%%%%%%%%%%%%%%%%%%%%%%%%%
\subsubsection{Partition Group Fitness}
\label{sssec:pg_fit}
The model is optimized by its fitness (power or throughput) as specified by the user. The partition group fitness ($PGF$) is calculated by summing up all of its partition's fitness. Since each partition is a sub-model which is mapped fully on chip, we can use previous optimization methods~\cite{ankit2019puma, sun2023pimcomp} to optimize a partition. We modify PIMCOMP~\cite{sun2023pimcomp}'s scheme for partition optimization by considering layer dependence and memory accesses as described in Sec.~\ref{ssec:partition_gen}.

% IS THIS TRUE? previous works that consider latency or throughput 
% In the power-optimized mode, we use the inter-layer pipelined execution scheme, to minimize the number of global memory access. In the throughput-optimized mode, we use the batch pipelined execution scheme.

%%%%%%%%%%%%%%%%%%%%%%%%%%%%%%%%%%%%%%%%%%%%%%%%%%%%%%%%%%%%%%%%%%%%%%%%%%%%%
\subsubsection{Partition Score and Selection}
\label{sssec:p_score}

We define a partition score to evaluate the performance of a partition relative to the overall population. Worse performing partition or a partition pair is selected for mutation. For a partition $P = \{x_i \mid a \leq i < b\}$, the partition score $R$ is defined as follows.
%We define the partition score which is used to calculate the relative performance of the selected partition. The relative performance $R_k[a,b]$ is defined over a span of partition units $\{x_i \mid a \leq i < b\}$ as follows.
\vspace{-1mm}
\begin{displaymath}
  \begin{gathered}
    m(x_i) = \frac{f(P)}{|P|} \quad \text{where } x_i \in P, \\
    \overline{F}[p,q] = \mathop{\mathbb{E}} \left[ \sum_{i=p}^{q-1} m(x_i) \right], \;
    R = \frac{ f(P) }{ \overline{F}[a,b] }
    \end{gathered}
\end{displaymath}

First, the partition unit fitness, $m(x)$, is defined as its residing partition's fitness ($f(P)$) divided by the number of partition units in the partition ($|P|$). This is further used to describe the expected fitness of the partition units' span, $\overline{F}[p,q]$. For each partition unit in the span $[p, q]$, their fitness is summed up and the expectation over the population $\Pi$. The partition score $R$ is given as the ratio between partition's fitness and the expected fitness over the same span. 

The score effectively captures whether the selected partition of an individual is performing well or not compared to other individuals in the pool. If there exists a partition which could potentially perform better if partition units are better partitioned, the $m(x)$'s would be relatively larger than other individuals where the same partition units reside in better performing partitions. Therefore, optimizing against the defined partition score would provide pressure to partition better.

\subsubsection{Mutation}
\label{sssec:mutation}
For $n_{mut}$ partition groups in each generation, single or a pair of partitions are selected according to the partition score and are mutated with one of the four mutation schemes. \textit{Merge} selects two neighboring partitions and merges them into a single partition. We evaluate the relative partition score of consecutive partition pairs to select the worst-performing pair to perform the merge. This method effectively removes small partitions that are inefficient. \textit{Split} transforms a selected partition into two randomly split partitions. This method removes ill-performing partitions holding many layers, suffering from low replication value.
\textit{Move} moves a partition unit between two neighboring partitions. This method adjusts the total fitness in a meticulous way, by searching for an optimal partitioning position for neighboring partitions.
\textit{FixedRandom} fixes a partition with best fitness and all other partitions are randomly generated. This guarantees that new individuals for the next generation are highly random and do not fall into a local optimum.

% The first method, \textit{Merge}, selects two neighboring partitions and merges them into a single partition. We evaluate the relative partition score of consecutive partition pairs to select the worst-performing pair to perform the merge. This method effectively removes small partitions that are inefficient. The second method, \textit{Split} transforms a selected partition into two randomly split partitions. This method removes ill-performing partitions holding many layers, suffering from low replication value. The third method, \textit{Move}, moves a partition unit between two neighboring partitions. This method adjusts the total fitness in a meticulous way, by searching for an optimal partitioning position for neighboring partitions. The final method is \textit{FixedRandom}, where a superior partition is fixed and all other partitions are randomly generated. This guarantees that new individuals for the next generation are highly random and do not fall into a local optimum. The intra-partition optimization is run for all newly generated partitions and forms mutated chromosomes for the next generation.

%%%%%%%%%%%%%%%%%%%%%%%%%%%%%%%%%%%%%%%%%%%%%%%%%%%%%%%%%%%%%%%%%%%%%%%%%%%%%
%%%%%%%%%%%%%%%%%%%%%%%%%%%%%%%%%%%%%%%%%%%%%%%%%%%%%%%%%%%%%%%%%%%%%%%%%%%%%
\section{Evaluation}
\subsection{Experiment Setup}
%%%%%%%%%%%%%%%%%%%%%%%%%%%%%%%%%%%%%%%%%%%%%%%%%%%%%%%%%%%%%%%%%%%%%%%%%%%%%
\subsubsection{Hardware Details}

\setlength{\aboverulesep}{0.5pt}  % toprule 위의 여백 설정
\setlength{\belowrulesep}{0.5pt}  % toprule 아래의 여백 설정

\begin{table}
\caption{Hardware Configuration}
\vspace*{-2mm}
\label{tab:hw_config}
\centering
\begin{tabular}{cccc}
\toprule
{  Component  } & { Parameters } & { Specification } & { Power($mW$) } \\
\midrule
VFU             & \# per core  & 12             & 22.8 \\
Local Memory    & \# per core  & 64kB           & 18.0 \\
Control Unit    & \# per core  & -              & 8.0 \\
DRAM            & config. & LPDDR3 8GB  & trace-based \\
\specialrule{0.8pt}{0pt}{3pt}
\end{tabular}
\begin{tabular}{ccccc}
\toprule
{ Chip } & { \# Cores } & { \# Crossbar/Core } & { Capacity(MB)} & { Power($W$) } \\
\midrule
S        & 16           & 9                    & 1.125           & 1.57 \\
M        & 16           & 16                   & 2.0             & 2.80 \\
L        & 36           & 16                   & 4.5             & 6.30 \\
\bottomrule
\end{tabular}
\end{table}

\begin{table}
\caption{Network Model and Compiler Support}
\vspace*{-2mm}
\label{tab:network}
\centering
\begin{booktabs}{cccc}
\toprule
{  Network  } & { Linear.(MB) }  & { Conv.(MB) } & { Total(MB) } & { Prev. } & { \textbf{Ours} }\\
\midrule
VGG16         & 58.95  & 7.02    & 65.97         & \xmark & \bcmark \\
ResNet18      & 0.244  & 5.324   & 5.569         & \xmark & \bcmark \\
SqueezeNet    & 0.0    & 0.58725 & 0.58725       & \cmark & \bcmark \\
\bottomrule
\end{booktabs}
\end{table}
\begin{figure*}[t]

    \centering
    \includegraphics[width=1.0\linewidth]{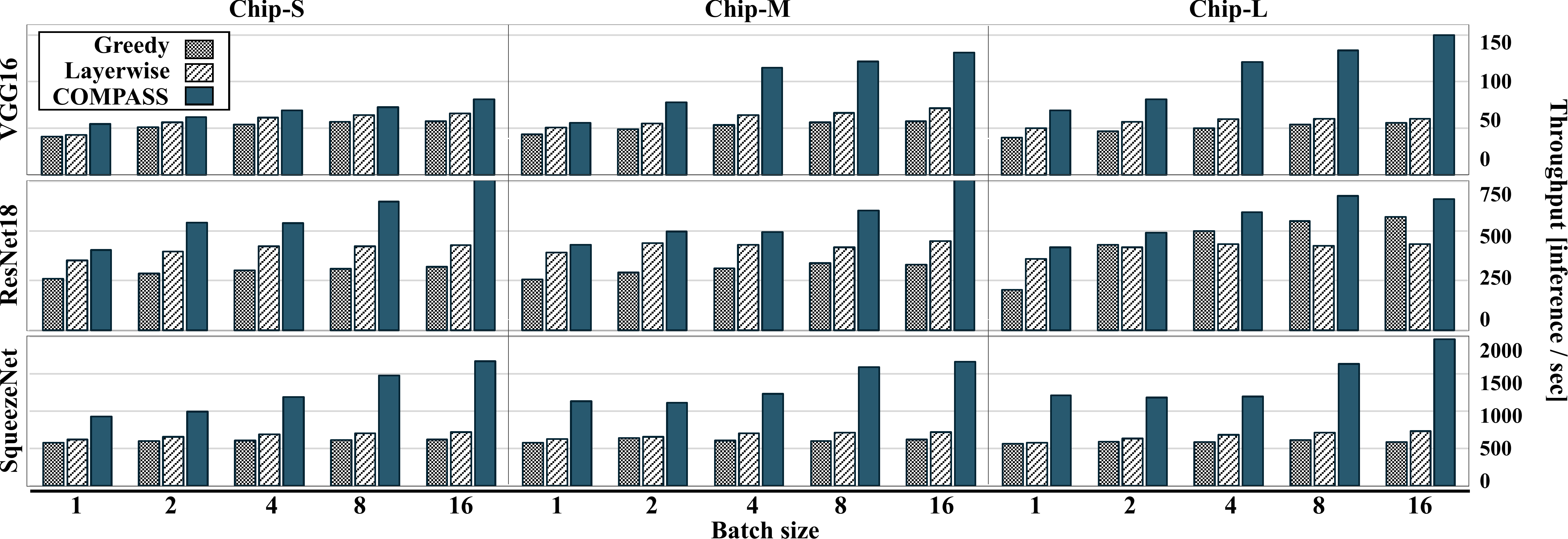}
    \vspace*{-7mm}
    \caption{Throughput comparison}
    \vspace*{-5mm}
    \label{fig:throughput}
\end{figure*}
We adopt the hardware architecture of PIMCOMP~\cite{sun2023pimcomp} and PUMA~\cite{ankit2019puma} but with tighter resource constraints. We adopt the parameters used in PIMCOMP for core design and scale them into 16nm technology, including the power information of VFUs, control units, data and instruction memory. We use a bus interconnect to connect the PIM cores. We adopt a 256 x 256 crossbar array and calculate the power consumption using the energy breakdown from the 16nm IMC-SRAM prototype by Jia et al.~\cite{jia202115}. The write power is directly taken from the prototype. The inference power is estimated by adding the ADC power and the power of remaining components which are scaled with respect to the number of wordlines. We model the DRAM energy by generating a memory trace from the scheduled instruction and feeding it into DRAMsim3~\cite{li2020dramsim3}. To demonstrate the generality of our approach, we evaluate on three chip configurations, ``S," ``M," and ``L," with varying memory footprint sizes. These sizes are chosen based on the fact that existing chip prototypes across various technologies~\cite{jia202115,sun2023pimcomp,xue202116,huang2023nonvolatile,hung20228,cai202333}, typically exhibit an in-memory footprint up to a few megabytes or less.The hardware configurations are summarized in Table~\ref{tab:hw_config}.

%%%%%%%%%%%%%%%%%%%%%%%%%%%%%%%%%%%%%%%%%%%%%%%%%%%%%%%%%%%%%%%%%%%%%%%%%%%%%
\subsubsection{Benchmark and Baselines}
We evaluate three representative CNN networks with varying model sizes: VGG16, ResNet18, and SqueezeNet. The parameter sizes of the networks are given in Table~\ref{tab:network}. We assume 4b weight and activation precision to faithfully model power consumption based on a recent CIM array which incorporates 4b quantization~\cite{jia202115}. We evaluate the networks across various chip configurations and batch sizes, labeling each evaluation as \textit{``{Network}-{ChipConfig}-{BatchSize}"} (e.g. \textit{``ResNet18-S-4"}). Note that existing compiler methods can only map SqueezeNet in resource-constrained chips, while \textit{COMPASS} allows all three models.

%with_write_end1/resnet18_c16x9_GA_element_latency_ppp_core_1/PartitionResult.txt

We compare \textit{COMPASS} results with two baseline partitioning schemes: \textit{greedy}, and \textit{layerwise}. The \textit{greedy} scheme attempts to pack in as many consecutive layers as possible, by iterating the nodes and tracking the remaining memory footprint. The \textit{layerwise} scheme maps a single Conv/Linear layer at a time, then maps the trailing non-Conv/Linear nodes together with its producer Conv/Linear nodes. All partitioning schemes, including ours, are implemented by extending the open-source PIMCOMP framework~\cite{sun2023pimcomp}. We implement code for the \textit{COMPASS} algorithm with an enhancement to PIMCOMP's latency estimator, as the original estimator was designed for non-partitioned design and did not consider weight load, intermediate data load/store.

%Table.~\ref{tab:partition_example} illustrates an example of the first 3 partition results on a \textit{``ResNet18-S-1"} configuration with different partitioning schemes. PS and PE indicate the start and end index of the partition unit inside each partition. The number of mapped nodes and the sum of all replication numbers in each partition are also shown. The model is divided into 60 partition units, and the first three partitions cover 100\%, 5\%, and 21.6\% of them in \textit{greedy}, \textit{layerwise}, \textit{COMPASS} respectively. We can see from the result that \textit{COMPASS} balances the trade-off between replication and layer-pipeline parallelism.

%%%%%%%%%%%%%%%%%%%%%%%%%%%%%%%%%%%%%%%%%%%%%%%%%%%%%%%%%%%%%%%%%%%%%%%%%%%%%
\subsubsection{\textit{COMPASS} GA algorithm Parameters}
We maintain a population of 100 for 30 generations, using a selection size ($n_{sel}$) of 20 and a mutation size ($n_{mut}$) of 80. We also adopt an early stopping mechanism. The mutation schemes are selected with the same probability. %If the result of a mutation does not satisfy the resource constraints, another mutation scheme is attempted, selected with equal probability.
% add description of GA

%%%%%%%%%%%%%%%%%%%%%%%%%%%%%%%%%%%%%%%%%%%%%%%%%%%%%%%%%%%%%%%%%%%%%%%%%%%%%
%%%%%%%%%%%%%%%%%%%%%%%%%%%%%%%%%%%%%%%%%%%%%%%%%%%%%%%%%%%%%%%%%%%%%%%%%%%%%
\subsection{Experimental Results}

\begin{figure}[t]
\vspace*{1mm}
    \centering
    \includegraphics[width=1.0\linewidth]{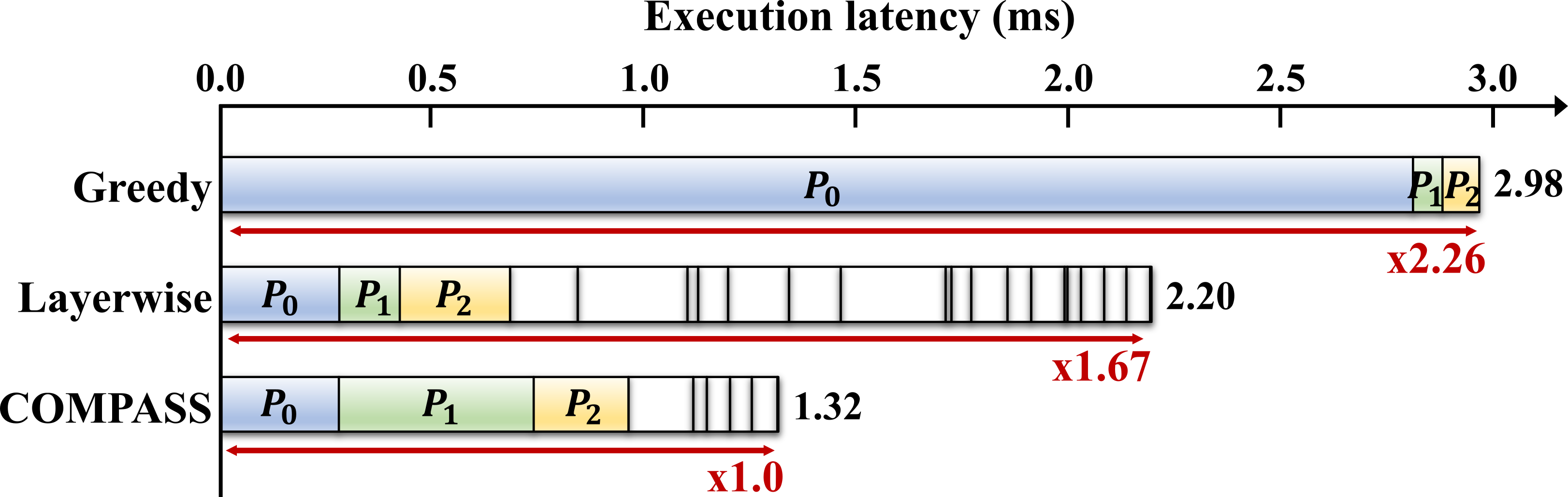}
    \vspace*{-7mm}
    \caption{Latency breakdown for each partition. The per-partition latencies for each schemes are shown, with the first three partitions highlighted for easier comparison.}
    \vspace*{0mm}
    \label{fig:per_partition_latency}
\end{figure}

%%%%%%%%%%%%%%%%%%%%%%%%%%%%%%%%%%%%%%%%%%%%%%%%%%%%%%%%%%%%%%%%%%%%%%%%%%%%%
\subsubsection{Latency and Throughput}
Fig.~\ref{fig:throughput} shows the result of inference throughput of \textit{COMPASS} under different workloads and memory constraints. \textit{COMPASS} achieves 1.78X higher throughput than the baseline methods. Specifically, \textit{COMPASS} outperforms the \textit{greedy} scheme by 1.80X, 1.71X, and 2.24X and \textit{layerwise} scheme by 1.56X, 1.31X, and 1.98X in VGG16, ResNet, and SqueezeNet respectively.

We can observe different trends in performance gain with differing chip sizes and networks. Performance gain in VGG execution for a large chip is much higher than the one for a smaller chip. This is due to the fact the VGG network has large channel dimensions, making it hard to fit in many layers together in a small chip. In such a case, possible pipeline depth and replication numbers reduce, thereby making both greedy and layerwise method perform similar. The optimal balance between pipeline and replication does not differ much either. On the other hand, a larger chip size can have bigger chances of tweaking replication numbers and the pipeline depth of each partition. ResNet18 and SqueezeNet models have relatively smaller layers and a small chip configuration is enough to exploit \textit{COMPASS}'s optimization ability.
In case of ``ResNet18-L", we see that all methods suffer a certain amount of performance degradation compared to smaller chip configuration. 
%We suspect that this is due to the fact that heuristic used for core mapping in PIMCOMP intra-partition optimization is sub-optimal. 
%The performance gain is relatively smaller in MobileNet due to large memory footprint. Since our method fails to replicate large Conv/Linear layers due to limited chip size, it does not gain additional benefit compared to other methods. 
 As we increase the batch size, the weights are written once and reused over multiple samples, effectively making throughput higher. Our methods outperform the baselines across all typical batch sizes.

Fig.~\ref{fig:per_partition_latency} shows the latency results for each partition during execution of \textit{``ResNet18-M-16"}. Different colored portion of the graph indicates different partition's execution time. While \textit{COMPASS} achieves 2.26X and 1.67X speed-up compared to greedy and layerwise partitioning respectively.
%we can also observe that execution latencies are more uniform in our method, indicating a more balanced partitioning result.

\textit{Greedy} partitioning maps many layers in the earlier partition and does not exploit weight replication favorably, resulting in high latency. Its first partition, $P_0$, occupies over 95\% of the total execution time. \textit{Layerwise} partitioning maps a single Conv/Linear layer on a partition and exploits more weight replications. However, this increases DRAM access, as more partitions require intermediate features to be moved in and out of DRAM between each partition. In contrast, \textit{COMPASS} can map multiple layers within a single partition, reducing this overhead. %This results in high overhead in final layers as the other two methods utilize layer-pipelining and do not require the movement of intermediate features.

%%%%%%%%%%%%%%%%%%%%%%%%%%%%%%%%%%%%%%%%%%%%%%%%%%%%%%%%%%%%%%%%%%%%%%%%%%%%%
\subsubsection{Energy-Delay-Product}

\begin{figure}[t]
    \centering
    \includegraphics[width=1.0\linewidth]{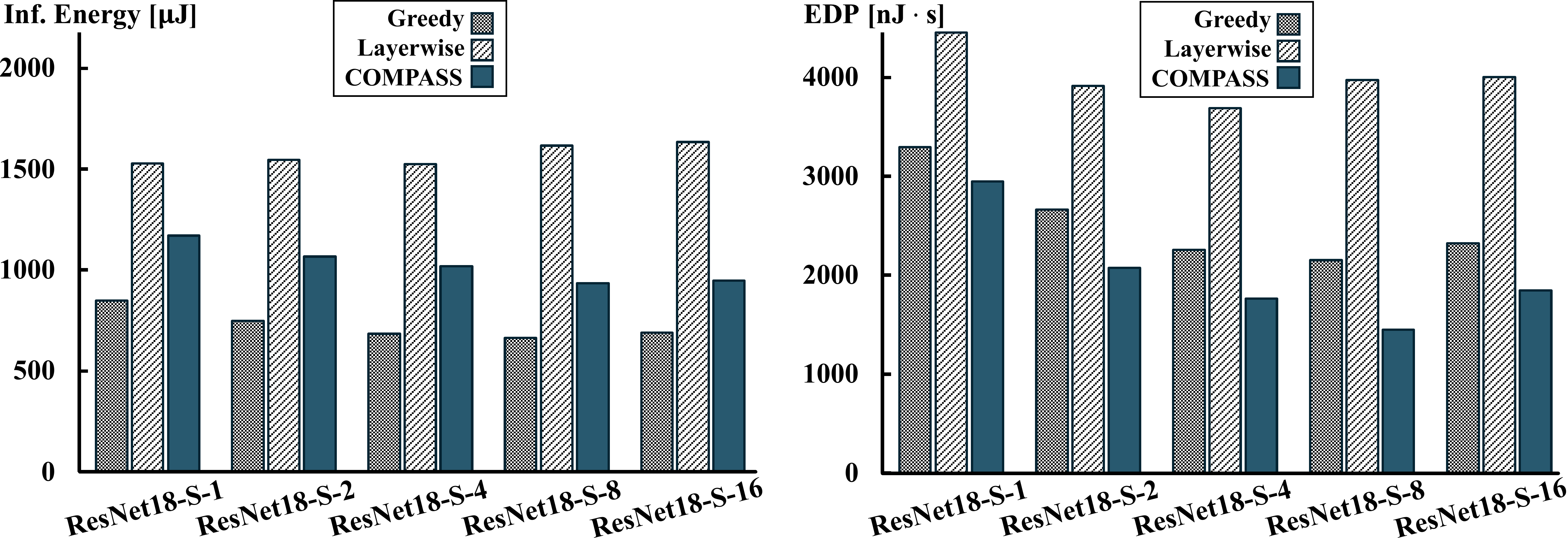}
    \vspace*{-7mm}
    \caption{Inference energy and energy-delay product}
    \label{fig:edp}
\end{figure}

Fig.~\ref{fig:edp} shows the inference energy and energy-delay product (EDP) per sample of \textit{COMPASS} for different batch sizes in a \textit{``ResNet18-S"} configuration. Since we optimize for latency, we utilize more replication in each partition, thereby limiting the pipeline depth compared to \textit{greedy} partition scheme. This requires more data communication with DRAM, increasing the energy consumption required per inference as shown in Fig.~\ref{fig:edp}. However, we observe that \textit{COMPASS} is more efficient in terms of joint optimization of power and latency. We outperform \textit{greedy} and \textit{layerwise} schemes by 1.28X and 2.08X in EDP results on average. This is mainly because limited replication in \textit{greedy} scheme incurs many stalls during the execution making it less optimized for latency.

%%%%%%%%%%%%%%%%%%%%%%%%%%%%%%%%%%%%%%%%%%%%%%%%%%%%%%%%%%%%%%%%%%%%%%%%%%%%%
\subsubsection{Effect of different batch sizes}

\begin{figure}[t]
    \centering
    \vspace*{0mm}
    \includegraphics[width=0.95\linewidth]{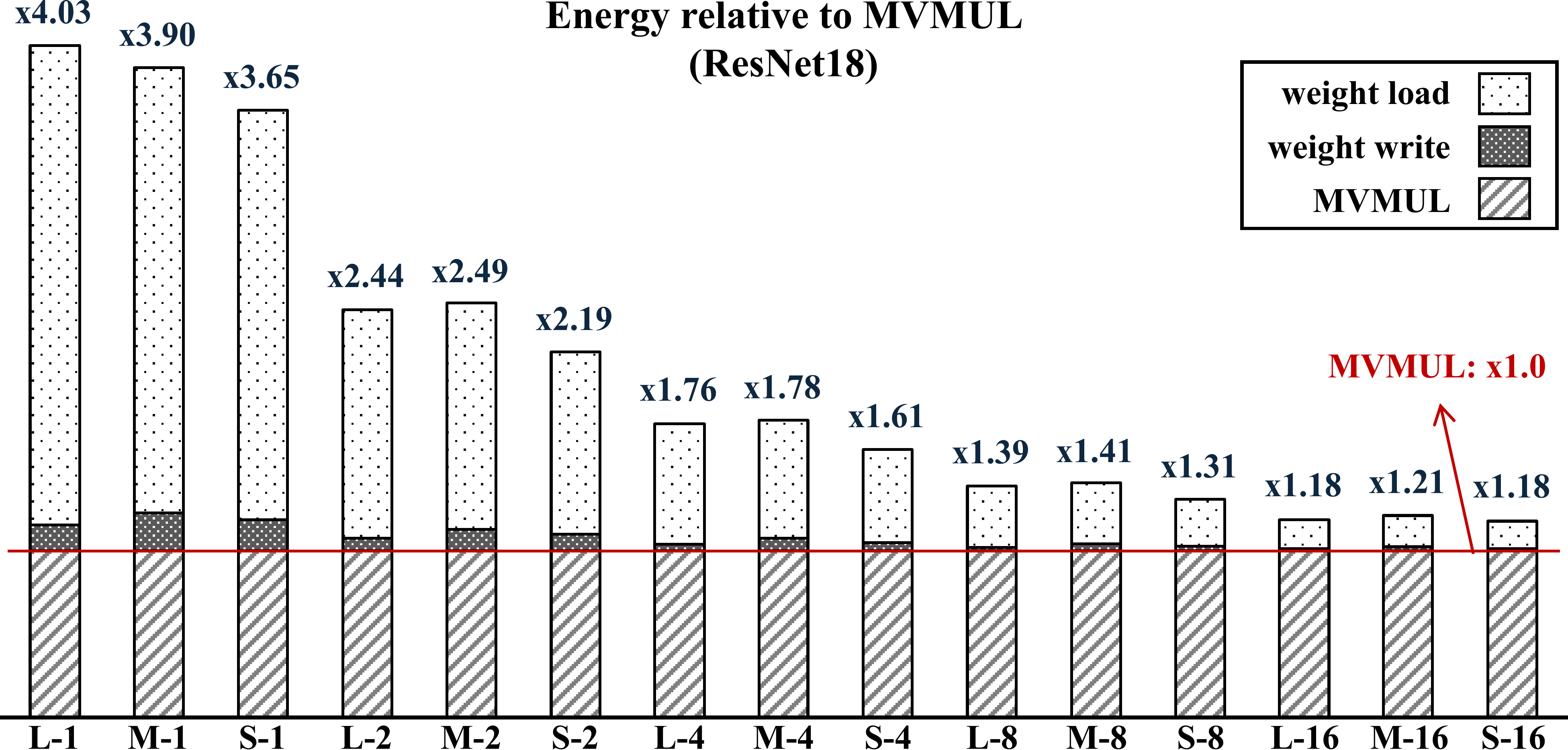}
    \vspace*{-2mm}
    \caption{Energy consumption of weight writes and loads relative to MVMUL for different chip and batch sizes. For example, ``M-4" denotes chip ``M" with a batch size of 4.}
    \label{fig:batch_size_effectiveness}
\end{figure}

As discussed in Sec~\ref{ssec:partitioned_exec}, an appropriate number of batch size is important to amortize weight replacement overhead. Fig.~\ref{fig:batch_size_effectiveness} plots the energy consumption of the weight writes and loads on different batch sizes normalized to matrix vector multiplication energy consumption. With a batch size of 1, the weight load energy dominates over compute energy. With a batch size of 16, replacement overhead is sufficiently amortized. 

%%%%%%%%%%%%%%%%%%%%%%%%%%%%%%%%%%%%%%%%%%%%%%%%%%%%%%%%%%%%%%%%%%%%%%%%%%%%%
\subsubsection{GA Fitness Convergence}

\begin{figure}[t]
    \centering
    \includegraphics[width=0.95\linewidth]{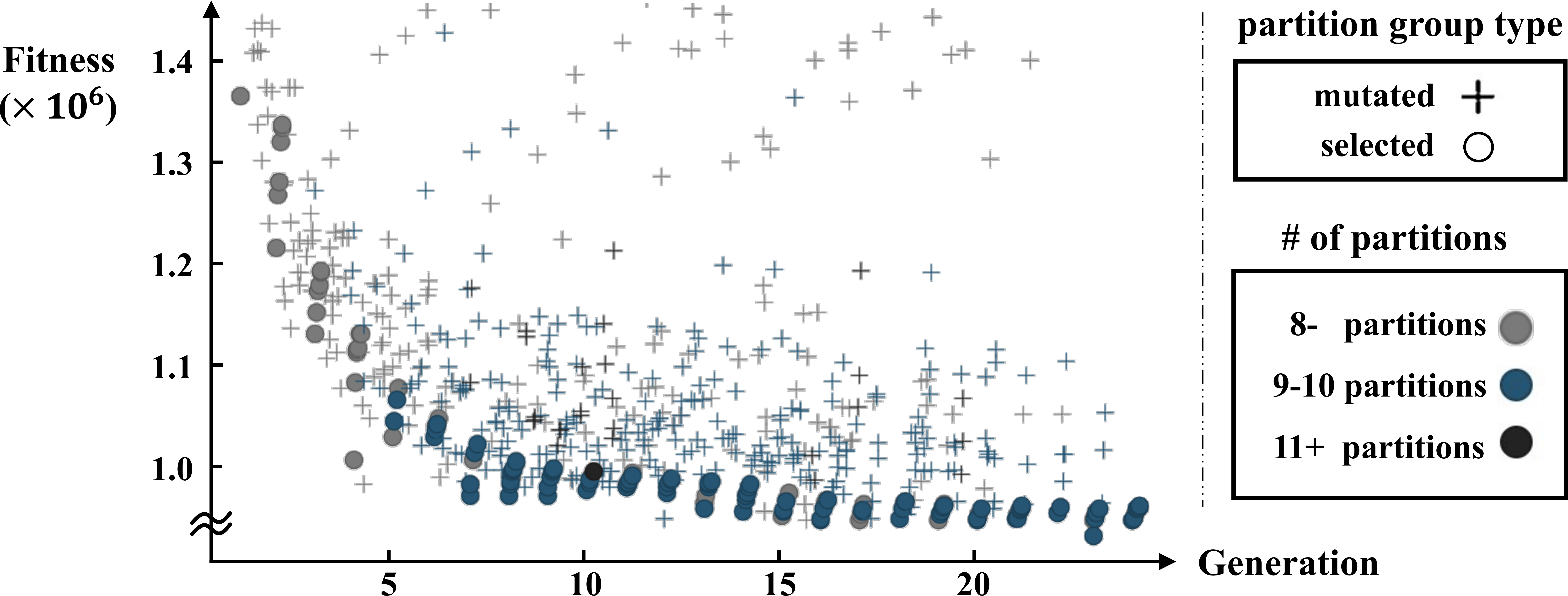}
    \vspace*{-3mm}
    \caption{Evolution of partition groups and their number of partitions over generations}
    \label{fig:ga_fit_progress}
\end{figure}

Fig.~\ref{fig:ga_fit_progress} shows the evolution of the population's fitness across generations under a \textit{COMPASS} \textit{``ResNet18-M-16"} optimization. For clarity, a random one-third of the population is selected for visualization in each generation. The population selected for the next generation ($\Pi_{sel}$)  is represented by ``O" markers, while the mutated population ($\Pi_{mut}$) is represented by ``+" markers. We can observe that GA algorithm makes the population steadily evolve into the selected one, finding an optimal fit.

The different partition groups are represented by various colors based on the number of partitions. Initially, the population tends to start with fewer partitions, and by the 9th or 10th generation, an optimal number of partitions is typically reached. From this point, most partition groups continue to be refined within the same partition count, yielding improved fitness as the GA algorithm explores the design space.
%Then, it optimizes within the same partitions and generates a better fit, as our GA algorithm navigates through the vast design space.

% Well-performing individuals are kept for the next generation, while mutation is applied to ill-performing partition within the individual. We can see in the figure that the initial generations begin with a relatively smaller number of partitions (less than 8) and evolves to find the optimal partition number, typically around 9 or 10. Note that even with the same number of partitions, we may still observe individuals with poor performances, as our genetic algorithm navigates through the vast design space.

%%%%%%%%%%%%%%%%%%%%%%%%%%%%%%%%%%%%%%%%%%%%%%%%%%%%%%%%%%%%%%%%%%%%%%%%%%%%%
%%%%%%%%%%%%%%%%%%%%%%%%%%%%%%%%%%%%%%%%%%%%%%%%%%%%%%%%%%%%%%%%%%%%%%%%%%%%%
\section{Discussion}
\subsection{Challenges of Adopting Traditional Compiling Methods for Digital GPU/NPUs to PIM Architectures}

While both Processing-In-Memory (PIM) cores and digital processing engines, such as GPU SM cores and other NPUs, serve as fundamental processing units for AI accelerations, their operational paradigms differ significantly. PIM cores perform computations directly where the data is stored, reducing data movement but incurring higher write costs. Also, data movement of PIM cores are done in crossbar granularity. In contrast, GPU SM cores and digital accelerators rely on fast data movement between the cores and the local buffer. These differences create challenges when applying traditional compiling methods to PIM architectures. This involves managing high memory write costs and ensuring efficient partitioning and scheduling within PIM's limited resources. As a result, new compilation techniques are needed to fully exploit PIM's unique capabilities while overcoming these constraints.

%%%%%%%%%%%%%%%%%%%%%%%%%%%%%%%%%%%%%%%%%%%%%%%%%%%%%%%%%%%%%%%%%%%%%%%%%%%%%
%%%%%%%%%%%%%%%%%%%%%%%%%%%%%%%%%%%%%%%%%%%%%%%%%%%%%%%%%%%%%%%%%%%%%%%%%%%%%
\subsection{Applicability to Different PIM Technologies}
While the current work is evaluated on an SRAM-based architecture, this choice is due to the maturity of in-memory SRAM technology, making it an easier target for system-level evaluation. However, our approach can also be extended to emerging non-volatile memory (eNVM) technologies such as ReRAM and MRAM. Although ReRAM is limited by its write endurance, our method aligns well with this constraint by minimizing the number of weight rewrites. In the case of MRAM, which has higher write latency and energy consumption, we can parameterize the crossbar properties as part of the hardware configuration and optimize weight replacement accordingly.

%%%%%%%%%%%%%%%%%%%%%%%%%%%%%%%%%%%%%%%%%%%%%%%%%%%%%%%%%%%%%%%%%%%%%%%%%%%%%
%%%%%%%%%%%%%%%%%%%%%%%%%%%%%%%%%%%%%%%%%%%%%%%%%%%%%%%%%%%%%%%%%%%%%%%%%%%%%
\section{Conclusion}
This work develops \textit{COMPASS}, a compiler framework for PIM-based CNN accelerator in which a given model does not fit entirely on chip. \textit{COMPASS} generates an optimal model partition where each partition fits on chip, thereby enabling automatic execution of larger neural networks without the need for manual model decomposition. 
To the best of our knowledge, \textit{COMPASS} is the first compiler framework to consider communication with external memory for analog in-memory computing hardware.
%COMPASS$ bridges the gap between the ReRAM-based PIM compilers and the SRAM-based PIM accelerators in practice.
Compared to naive partitioning schemes, \textit{COMPASS}'s partitioning scheme achieves higher throughput and better EDP in diverse workload settings. 

% \begin{acks}
% \end{acks}

\bibliographystyle{IEEEtran}
\bibliography{refbib}

\end{document}